\documentclass[twocolumn]{aastex63}

\usepackage{amsmath}

\accepted{May, 2021}

\submitjournal{ApJ}

\shorttitle{The Escapers}
\shortauthors{Abbas et al.}

\graphicspath{{./}{figures/}}

\begin{document}

\title{RR Lyrae Stars In Stellar Streams with Gaia: The Escapers}

\email{moe.abbas@nyu.edu}

\author[0000-0001-9146-0421]{Mohamad Abbas}

\affiliation{Center for Astro, Particle, and Planetary Physics (CAP$^3$), New York University Abu Dhabi}
\affiliation{New York University Abu Dhabi, PO Box 129188, Saadiyat Island, Abu Dhabi, United Arab Emirates}

\author{Eva K. Grebel}
\affiliation{Astronomisches Rechen-Institut, Zentrum f\"{u}r Astronomie der Universit\"{a}t Heidelberg, M\"{o}nchhofstr. 12--14, D-69120 Heidelberg, Germany}

\author{Mirko Simunovic}
\affiliation{Subaru Telescope, 650 N Aohoku Pl, Hilo, HI 96720, USA}

\begin{abstract}

We attempt to identify RR Lyrae (RRL) stars in stellar streams that might have escaped from seven globular clusters (GCs) based on proper motions, distances, color-magnitude diagrams, and other properties extracted from the Gaia Early Data Release 3 (EDR3) database. Specifically, we cross-match two large RRL stars catalogs (from Gaia DR2 and Catalina Sky Survey) with each other and with the EDR3 database and we end up with a sample of $\sim$ 150,000 unique RRL stars. We calculate distances to RRL stars using the $M_{G}$ -- [Fe/H] and $M_{V}$ -- [Fe/H] absolute magnitude-metallicity relations and adopt [Fe/H] values for the GCs from different spectroscopic studies. We also constrain our search in areas where stellar streams associated with GCs were previously suggested or identified in other studies. We identify 24 RRL stars that might have escaped from the following seven GCs: Palomar 13 (Pal 13), NGC 6341 (M92), NGC 5904 (M5), NGC 5466, NGC 1261, NGC 288, and NGC 1851. We list our findings in Table \ref{TableEscapers}. 

\end{abstract}


\section{Introduction}\label{sec:intro}
The Milky Way Galaxy is one of the most studied galaxies in the universe and yet its formation history, evolution, dynamics,  and dark matter properties are not fully understood. Over the past decades, however, large-sky surveys allowed astronomers to map different parts of the Galaxy in great detail. This revealed many hidden facts about some historic events that occurred over the past billions of years especially in its halo.  

For instance, various substructures, over-densities, and streams of stars were discovered during the two past decades. At first, the largest, densest, and most-prominent substructures were discovered \citep{ibata1994,grillmair1995,odenkirchen2001,newberg2003,bell2008,jordi2010} and as the large-sky surveys became more sensitive, more diffuse and less-dense substructures started to become unveiled, which helped us better understand the process of evolution of the Galaxy (e.g., \citealt{vivas2006, duffau2006, belokurov2007, carollo2007,sesar2010,martin2013, bernard2014,malhan2018a,shipp2018, Stringer2019}, and others). In a number of cases, some streams discovered at different times and locations were thought to be independent and to have different origins. With the help of follow-up observations and simulations, it became clear that some of these streams shared the same origins and progenitors.

In addition, different turning points in the halo of the Galaxy were detected using different methods. These turning points are associated with breaks in the power-laws of the halo and reflect different formation scenarios between the inner and outer halo (e.g., \citealt{bullock2005,johnston2008,das2016,drake2013a}). In brief, it is believed that the inner halo was mainly formed by the accretion of external massive systems and in situ star formation processes, while the outer halo was mainly formed from mergers of smaller and older systems (e.g., \citealt{delucia2008,schlaufman2009,font2011,beers2012,gomez2016}). The latter stars are usually old and metal-poor and studying them is of a great importance to understand the evolution of galaxies.

Stellar archaeology is thus crucial to study the properties and kinematics of halo stars and to trace back their origins. Identifying the origins of halo stars and stellar streams will significantly improve our understanding of galaxy evolution and will inform simulations. So far, more than 70 Milky Way stellar streams have been detected in the halo but not all of them have been intensively studied and in many cases their origins are still ambiguous \citep{mateu2018}.

As inferred above, some of the halo stars were formed inside the Galaxy while others were contributed by dwarf galaxies and globular clusters (GCs) that were accreted by the Milky Way. Tidal disruptions result in streams of stars in the vicinity of the accreted systems. In general, identifying which halo stars once belonged to larger accreted systems is challenging.

To confirm whether a stream of halo stars originated from a GC for example, the age, metallicities, and space velocities of the stars and GCs should be comparable. Given the number of stars at any line of sight in the halo, the thin and faint properties of some streams, and the several degrees on the sky that some streams can cover in the sky, this can be time consuming and not always feasible.

In the past few years, the Gaia database \citep{gaiadr1} was used to search for halo stars that escaped from dwarf galaxies and GCs to form stellar streams. 

For example, using the second data release of the Gaia mission (Gaia DR2; \citealt{gaiadr2}), \citet{thomas2020} and \citet{sollima2020} detected a stellar stream around NGC 6341 (M92) using different filtering techniques in addition to colors, magnitudes, parallaxes, and proper motion (PM) information of blue horizontal branch (BHB), main sequence (MS), and and red giant branch (RGB) stars. \citet{thomas2020} performed N-body simulations and suggested that the stream around M92 is relatively new and formed during the last $\sim$ 500 Myr.


\citet{shipp2020} studied tidal tails around GC Palomar 13 (hereafter, Pal 13) using photometric data from the Dark Energy Camera Legacy Survey (DECaLS\footnote{\url{https://www.legacysurvey.org/decamls/}}) of the DESI Legacy Imaging Surveys \citep{Dey2019}. They used a matched-filter technique and were able to detect tidal tails around Pal 13 that are aligned with the PM of Pal 13. Specifically, their findings were based on stars at the
blue edge of the MS as the latter stars provide high signal-to-noise to the matched-filter technique. Not only that, they also identified six RR Lyrae (RRL) stars of which three are located in the tails of Pal 13 and three located inside the GC. These findings allowed them to estimate the distance and total luminosity of the cluster.

\citet{hanke2020} used seven-dimensional chemodynamical data from
Gaia DR2 data and the Sloan Digital Sky Survey (SDSS, \citealt{york2000,sdssdr142018}) spectra and identified a large number of halo field
giants that likely originated in more than ten GCs. They used RRL stars
to estimate the distances to a number of the uncovered streams.

Moreover, \citet{vivas2020} identified RRL stars that belong to Ultra-Faint dwarf (UFD) galaxies using the Gaia DR2 catalog of RRL stars \citep{clementini2019}. Based on star locations, magnitudes, distances, and PMs, \citet{vivas2020} associated 47 RRL stars to 14 UFD galaxies (see table 1 and 2 from \citealt{vivas2020} for a complete list of their findings). 

\citet{pal5price2019} looked for RRL stars belonging to the Palomar 5 (hereafter, Pal 5) stellar stream \citep{odenkirchen2001}. They used the PanSTARRS-1 (PS1; \citealt{ps1,chambers2016}) catalog of RRL stars from \citet{sesar2017} and cross-matched it with Gaia DR2 data. They identified 27 RRL stars that are most likely members of the Pal 5 stream based on kinematic properties and estimated the cluster's distance to be 20.6 $\pm$ 0.2 kpc.

While this manuscript was in preparation, \citet{ibata2020} detected various halo streams using the Gaia DR2 and Early Data Release 3 (EDR3; \citealt{gaiaedr3,edr3_2,edr3_1}) catalogs. Specifically, they used the \emph{STREAMFINDER} algorithm \citep{malhan2018} and traced the sky positions of different streams.

In this paper, we search for RRL stars that might have escaped from seven GCs: Pal 13, NGC 6341 (hereafter, M92), NGC 5904 (hereafter, M5), NGC 5466, NGC 1261, NGC 288, and NGC 1851. We investigate the latter GCs as stellar streams associated with them were previously suggested or identified and we are currently working on a more in-depth analysis on other GCs (Abbas et al. 2021, in preparation).

RRL stars are intrinsically pulsating short-period variable stars with periods less than $\sim$ 1.3 days (d). These are core helium burning stars in the instability strip of the Hertzsprung–Russell diagram \citep{smith1995,layden1996}. There are several sub-types of RRL stars classified according to their periods, amplitudes, and shapes of their light-curves. The two most numerous sub-types are the RRc and RRab stars.

There are basically two ways how a GC can lose
stars. One is through internal dynamical effects (e.g., two-body
encounters), the other one due to the external gravitational field. Hence, detecting extratidal stars does not necessarily
imply that the GC is disrupting, though it might, especially when tidal
tails have been detected and if the RRL stars seem to be associated
with those tails. Identifying extratidal stars and their possible associations with GCs is thus important for a better understanding of the internal and external dynamical evolution of the GCs and can provide us with a a more comprehensive idea about the history of evolution of the Galaxy.

We specifically use RRL stars in our paper for several reasons. These stars have been used as powerful tools to investigate the accretion history of the Milky Way (e.g., \citealt{sesar2010,abbas2014b,drake2014,fabrizio2019,medina2020,prudil2021}). Some of them witnessed the early formation of galaxies so they act as fossils in stellar archaeology. Also, it is relatively easy to identify them based on the shapes of their light-curves as long as multi-epoch observations are available. Last but not least, RRL stars act as standard candles with their distance information imprinted in their properties. For instance, \citet{tatiana2018} derived a new absolute magnitude-metallicity relation in the Gaia G band ($M_{G}$ -- [Fe/H]) which we use in this study. 

In order to look for escapers, we use the Gaia DR2 catalog of RRL stars \citep{clementini2019} in addition to the full catalog of RRL stars from the Catalina Sky Survey (hereafter, the CRTS catalog; \citealt{drake2013a,drake2013b,drake2014,torrealba2015,drake2017}). We cross-matched both catalogs with each other and with the Gaia EDR3 database and ended up with $\sim$ 150,000 unique RRL stars. We then used the latter full catalog to search for RRL stars that might have originated from the seven GCs based on distances, PMs, color-magnitude diagrams (CMDs), and different statistical information. Our search was also based on the locations of some of the streams that we adopt from \citet{ibata2020} and \citet{thomas2020}.

This paper is organized as follows. The Gaia and CRTS surveys are briefly presented in Section \ref{sec:data} where the methods used to find the RRL stars are also discussed. In Section \ref{sec:method}, we describe our adopted method and selection cuts to find RRL stars that are most likely to have been diffused from our studied GCs. In the latter section, we also discuss the method used to find distances to the RRL stars and we review the completeness and purity levels of the catalog. In Section \ref{sec:analysis}, we briefly present the properties of each GC and apply our method to find possible RRL escapers. We also publish a list of RRL escapers in Table \ref{TableEscapers}. Finally, we summarize our work and results in Section \ref{sec:summary}.


\section{Data} \label{sec:data}
This study is based on two RRL stars catalogs from two sky surveys: Gaia and the CRTS. 

\subsection{The Gaia Survey} \label{sec:gaia}
Gaia was designed to be the largest precise astrometry mission and to multiply observe billions of sky objects over a five-year period. Gaia's catalogs \citep{gaiadr1,gaiadr2,gaiaedr3} hold a photomtetric and astrometric data set that contains the positions, parallaxes, and mean PMs for its targets. Most recently, the Gaia EDR3 was released on December 3rd, 2020. The new release has improved PMs and parallaxes information for more than 1.4 billion sources with an additional one year of observations compared to Gaia DR2.  

\subsubsection{RRL stars from Gaia} \label{sec:gaiarrl}

Using the Gaia DR2 multi-band time-series photometry, \citet{clementini2019} provide a catalog of $\sim$ 140,000 RRL stars of which $\sim$ 98,000 stars are of RRab type and the remaining $\sim$ 40,000 stars are of RRc type. 

The catalog of RRL stars from \citet{clementini2019} provides the Gaia identifier $ID_{Gaia}$, positions, average magnitudes in three wide bands ($\langle G \rangle$, $\langle G_{BP} \rangle$, and $\langle G_{RP} \rangle$), amplitudes in these different bands (A$_{G}$, A$_{BP}$, and A$_{RP}$), periods in days (P$_{G}$), the number of observations in different bands, and other information.

\subsection{The CRTS Survey} \label{sec:crts}

The CRTS used three different ground-based telescopes located in three different locations. All three telescopes are equipped with unfiltered 4k $\times$ 4k CCD cameras. The CRTS' main goal was to discover transient phenomena so it observed the sky multiple times. Its data was also used to investigate the variability of stars like RRL stars. On average, each RRL star was observed $\sim$ 200 times. 

\subsubsection{RRL stars from CRTS} \label{sec:crtsrrl}

In a series of papers, the full catalog of $\sim$ 42,000 RRL stars ($\sim$ 32,000 and 9,000 RRab and RRc stars, respectively)  discovered by the CRTS data was published \citep{drake2009,drake2013a,drake2013b,drake2014,torrealba2015}. The CRTS catalog of RRL stars includes the CRTS identifier $ID_{CRTS}$, positions, average magnitudes ($\langle V_{CRTS} \rangle$), amplitudes of variation (A$_{V}$), periods in days (P$_{V}$), the number of observations ($N_{CRTS}$), distances assuming an average absolute magnitude of $M_{v}$ = 0.61, and other information.

\section{The Method} \label{sec:method}
We use the combined catalogs from Gaia and CRTS described in Section \ref{sec:gaiarrl} and \ref{sec:crtsrrl}, respectively. We start by cross-matching both catalogs with EDR3 \citep{gaiaedr3} and with each other based on sky coordinates and using a 2$\arcsec$ radius.

The final full catalog contains $\sim$ 150,000 unique RRL stars. Around 75$\%$ of the RRL stars are found in the Gaia catalog but not in the CRTS catalog and 9$\%$ of them are found in the CRTS catalog but not in the Gaia catalog. The remaining 16$\%$ of the RRL stars are found in both catalogs. The Gaia survey covers a much larger portion of the sky compared to the CRTS survey and thus one would expect a larger number of RRL stars to be included in the Gaia catalog.


\subsection{Completeness and Purity Levels}\label{sec:completeness}

RRL stars in the final full catalog come from two different surveys that have different completeness and purity levels. The completeness level reflects the recovered fraction of RRL stars while the purity level reflects the fraction of true RRL stars in the catalog. The CRTS catalog has a completeness level of  $\textless$ 85$\%$ and purity level of $\textless$ 80$\%$ for objects closer than $\sim$ 25 kpc while the average completeness level for RRL stars from Gaia is $\sim$ 60$\%$--70$\%$ \citep{clementini2019}. Both, the completeness and purity levels decrease with increasing brightness and can vary from one area of the sky to the other. For instance, the completeness and purity levels of the Gaia RRL Catalog reaches $\sim$ 95$\%$ in the Magellanic clouds and Milky Way bulge regions and it drops to lower values in other parts of the sky.

Since we are studying RRL stars around relatively nearby GCs (7.5--26.0 kpc distance range), and since we targeted areas where the footprint of both RRL star catalogs overlap, we believe that the completeness and purity levels of our combined catalogs are at least or greater than the values of the CRTS catalog (i.e.,  completeness and purity levels $\textgreater$ 85$\%$ and $\textgreater$ 80$\%$, respectively). Additionally, we visually inspected phased light curves of all RRL stars passing our selection cuts using the CRTS and Gaia and we find that their variability classifications are highly reliable.

\subsection{Distances and PMs}\label{sec:distances} 
We adopt different approaches to find the final distances to the RRL stars (hereafter, $d_{RRL}$). 

For RRL stars that are exclusively found in the Gaia catalog, we use the absolute $G$ magnitude-metallicity relation ($M_{G}$ -- [Fe/H]) from \citet{tatiana2018} to calculate $d_{RRL}$. The distances for the RRL stars found in the CRTS catalog but not in the Gaia catalog are calculated using the absolute $V$ magnitude-metallicity relation ($M_{V}$ -- [Fe/H]) from \citet{drake2013a}. We did not use the distances provided in the CRTS catalogs as they were calculated using constant absolute magnitudes and metallicities. Finally, distances for RRL stars available in both catalogs were calculated two times using the $M_{G}$ -- [Fe/H] and $M_{V}$ -- [Fe/H] relations on the Gaia and CRTS data, respectively. In the latter cases, $d_{RRL}$ is considered to be average of the two calculated distances. The $M_{G}$ -- [Fe/H] and $M_{V}$ -- [Fe/H] adopted in this study were computed using RRab and RRc stars. However, since RRab stars have more distinctive light curves, and larger pulsation amplitudes and purity levels than RRc stars, the absolute magnitude-metallicity relations are usually more reliable when calculating distances to RRab stars. Nevertheless, these relations can and have been used to find distances to RRc stars as well. Finally, in order to find distances, we adopt reddening values from \citet{schlafly2011} dust maps to correct the magnitudes for extinction using the Python \emph{dustmaps} package \citep{pythondustmaps}.

Since the distances to RRL stars are calculated using the $G$ and $V$ absolute magnitude-metallicity relations that are sensitive to metallicities, we select regions around the GCs we are studying, adopt [Fe/H] of the selected GC, and use the $M_{G}$ -- [Fe/H] and $M_{V}$ -- [Fe/H] relations to find the distance to each RRL. Next, we adopt mean PMs for the GCs from \citet{eugene2019}. The different adopted values including distances to GCs (D$_{GC}$) are summarized in Table \ref{GCTableSummary}.



\begin{table*}
 \centering
  \caption{The Studied Globular Clusters}
  \begin{tabular}{cccccccc}
  \hline
  \hline
  GC & R.A.\footnote{Equatorial J2000.0 R.A. and Dec. are given in decimal degrees.\label{RaDec1}} (deg) & Dec.\textsuperscript{\ref{RaDec1}} (deg) & D$_{GC}$\footnote{\citet{eugene2019} and \citet{harris1996}.} (kpc) & ($\mu_{\alpha}$,$\mu_{\delta}$)\footnote{From \citet{eugene2019}, not corrected for the solar reflex motion.\label{GCTable}} mas.yr$^{-1}$ & err$_{(\mu_{\alpha},\mu_{\delta})}$\textsuperscript{\ref{GCTable}} mas.yr$^{-1}$& [Fe/H] & N$_{RRL}$\footnote{\smash{Number of RRL stars that might have escaped from the assigned GC. See Table \ref{TableEscapers} for more info.}}\\
  \hline
   Pal 13 & 346.685 & 12.772 & 26.0 & (1.615,0.142) & (0.101,0.089) & $-$1.91\footnote{\citet{koch2019}.} & 1--4 \\

   M92 (NGC 6341) & 259.281 & 43.136 & 8.3 & ($-$4.925,$-$0.536) & (0.052,0.052) & $-$2.35\footnote{\citet{carretta2009}.} & 1--2 \\

      M5 (NGC 5904) & 229.638 & 2.081 & 7.5 & (4.078,$-$9.854) & (0.047,0.047) & $-$1.29\footnote{\citet{harris1996}. \label{harris}} & 1--7 \\

       NGC 5466 & 211.364 & 28.534 & 16.0 & ($-$5.412,$-$0.8) & (0.053,0.053) & $-$1.98\textsuperscript{\ref{harris}}  & 1--5 \\

    NGC 1261 & 48.068 & -55.216 & 17.2\footnote{\citet{Ferro2019}\label{ferro}.} & (1.632,$-$2.038) & (0.057,0.057) & $-$1.42\textsuperscript{\ref{ferro}} & 1--5 \\

     NGC 288 & 13.188 & -26.583 & 8.9 & (4.267,$-$5.636) & (0.054,0.053) & $-$1.32\textsuperscript{\ref{harris}} & 1--5 \\
     
       NGC 1851 & 78.528 & $-$40.047 & 12.1 & (2.12,$-$0.589)  & (0.054,0.054) & $-$1.18\textsuperscript{\ref{harris}} & 1--6
      \\
  \hline
\end{tabular}
\label{GCTableSummary}
\end{table*}

\subsection{Selection Cuts}\label{sec:selectioncuts}
We start by comparing the properties of the GCs (i.e., distances and PMs) with the properties of RRL stars around them. Hereafter, all PM values are corrected for the solar reflex motion using the GALA package \citep{GalaPrice2017} unless stated otherwise. For simplicity, we will also remove the asterisk sign from $\langle \mu^{*}_{\alpha}\rangle$. First, we select all RRL stars around the GCs we are studying and apply the below selection cuts:

\begin{equation} \label{cut_dist}
D_{GC} - 0.2D_{GC} \;\; < \;\; d_{RRL} \;\; < \;\; D_{GC} + 0.2D_{GC}
\end{equation} 

\begin{equation} \label{cut_length}
|\mu_{RRL}| \;\;< \;\; 1.5|\mu_{GC}|
\end{equation} 

\begin{equation} \label{cut_angle}
|\theta_{GC} - \theta_{RRL}| \;\; < \;\; 15^\circ
\end{equation} 
\\

where $D_{GC}$ and $d_{RRL}$ are the distances for each GC and RRL star, respectively. $|\mu_{GC}|$ and $\theta_{GC}$ are the norm and angle of the corrected PM for each GC, respectively (and similarly for $|\mu_{RRL}|$ and $\theta_{RRL}$).

Equation \ref{cut_dist} gets rid of RRL stars that are too far from the selected GC while Equation \ref{cut_length} select stars with PM norm values that are maximum 1.5 that of the selected GC. In \citet{thomas2020} a value of 2 was adopted for the same selection cut but we decided to be more conservative to achieve a higher purity level. Finally, Equation \ref{cut_angle} remove stars that have PM directions that are moderately different from that of the selected GCs.

\begin{figure}
\centering
\hspace*{-0.41in}
\includegraphics[scale=0.33]{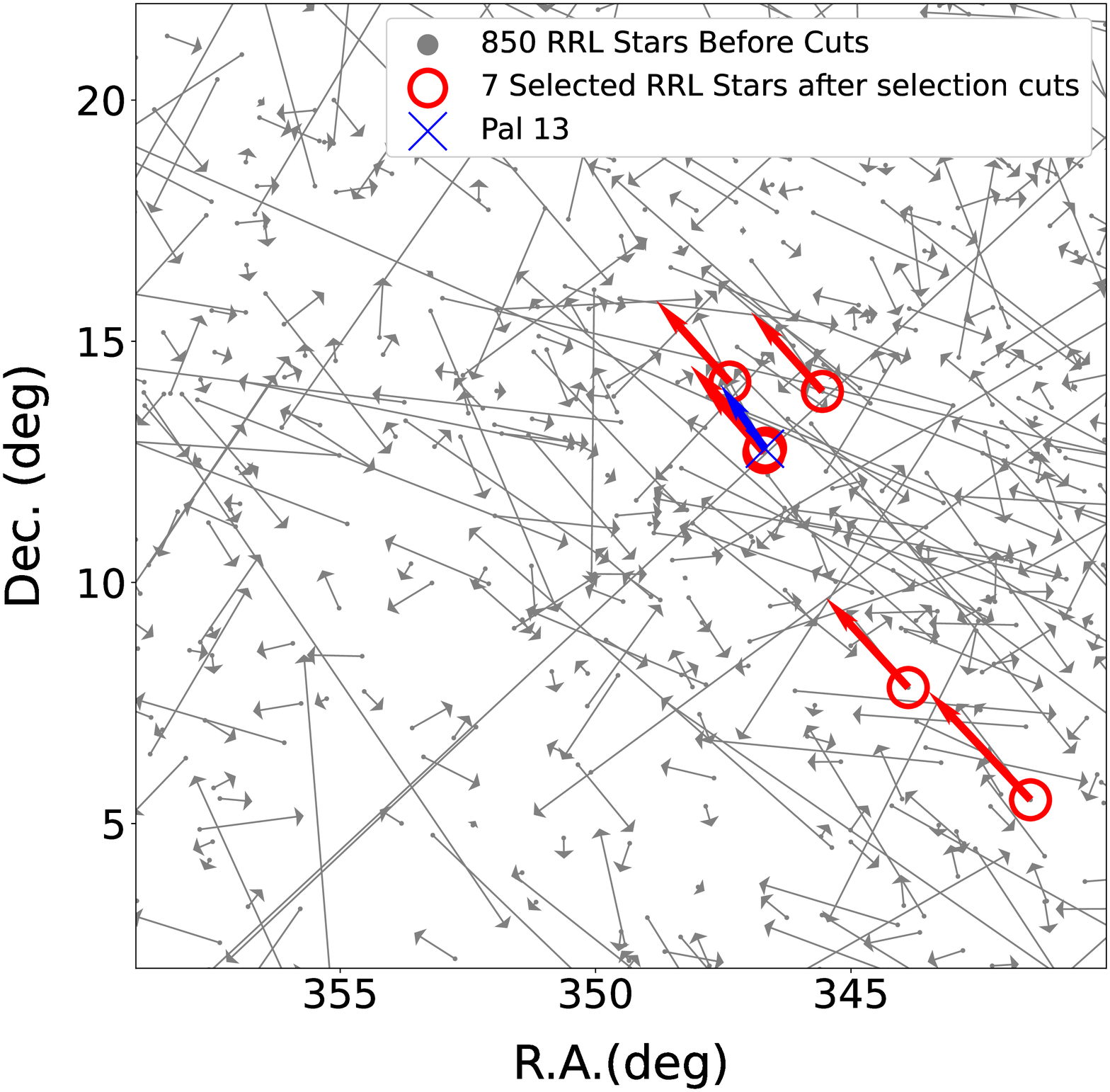}
\caption{850 RRL stars in the vicinity of Pal 13 and their PM arrows are shown in grey while the ones that passed the selection cuts (Equations \ref{cut_dist}, \ref{cut_length}, and \ref{cut_angle}) are shown in red open circles. Pal 13's location and PM arrow are plotted in a blue.}
\label{Pal13BeforeAfterCuts}
\end{figure}

As an example, the locations (R.A. and Dec. in degrees) and corrected PM directions of all RRL stars in the vicinity of Pal 13 are plotted in grey in Figure \ref{Pal13BeforeAfterCuts}. Stars that passed the first selection cuts (Equations \ref{cut_dist}, \ref{cut_length}, and \ref{cut_angle}) are shown with red open circles. Pal 13's location and PM arrow are shown in blue in the latter figure. These cuts reduced the number of RRL star candidates in the vicinity of Pal 13 from 850 to 7 stars only of which 3 are cluster members. 
\newline 

\section{analysis} \label{sec:analysis}
In this Section we investigate the properties of RRL stars around each of the seven GCs based on distances, PMs, light-curves, selection cuts, CMDs, and other properties.

We present our findings in Table \ref{TableEscapers} and we assign each RRL star a letter Group (A, B, or C). The letter Group reflects how confident we are about the origin of each RRL star. For instance, RRL stars assigned to Group A are very likely to have originated from the assigned GC. Stars selected with intermediate confidence are designated as Group B while those selected with less confidence are designated as Group C. The likelihood of stars being associated with the GCs (Group A, B, and C) is based on the locations of the candidate stars in different 2D plots and on the CMD diagram of each GC (e.g., see Section \ref{sec:pal13}).

\begin{figure*}
\centering
\hspace*{-0.7in}
   \includegraphics[scale=0.4]{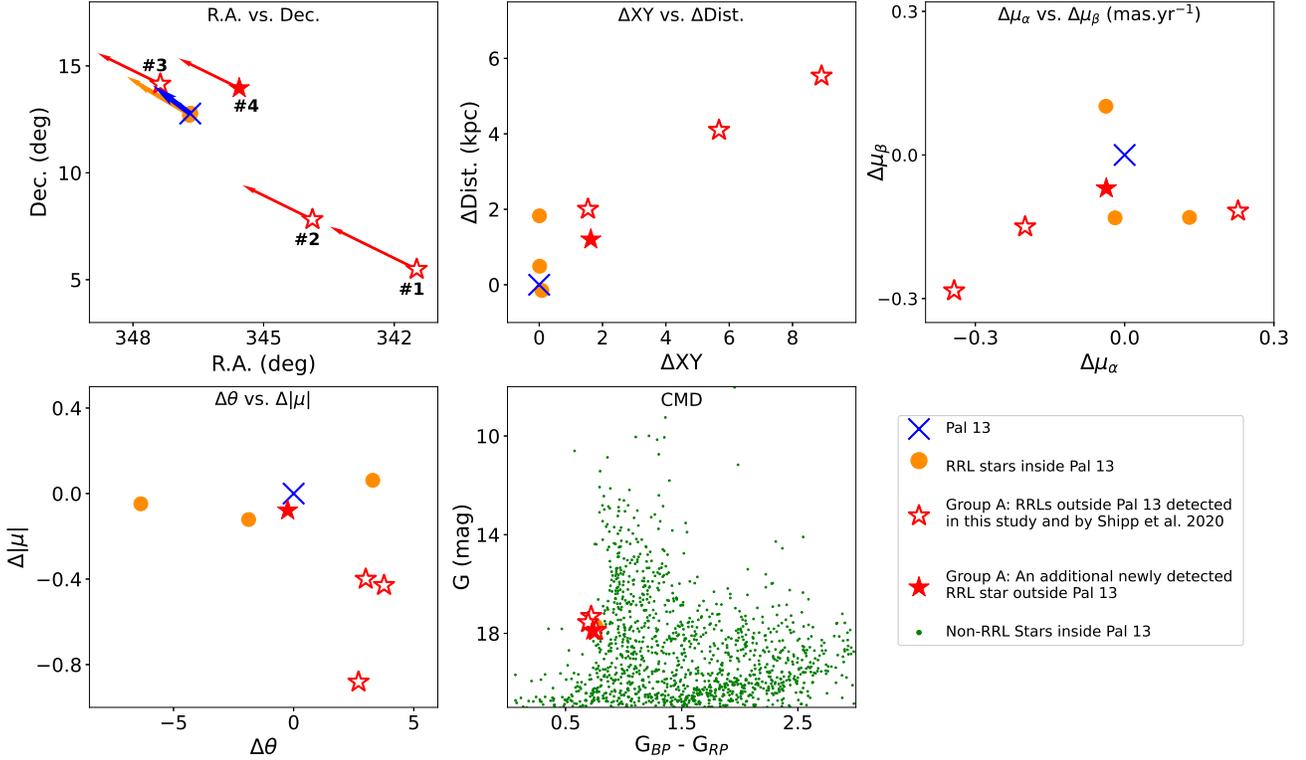}   
\caption{Different 2D plots such as (R.A. vs Dec.), ($\Delta$XY vs. $\Delta$Dist.\textsuperscript{\ref{DeltaDistFootNote}}), ($\Delta$$\mu_{\alpha}\textsuperscript{\ref{DeltaDistFootNote}}$ vs. $\Delta$$\mu_{\beta}$\textsuperscript{\ref{DeltaDistFootNote}}), and ($\Delta$$\theta$\textsuperscript{\ref{DeltaDistFootNote}} vs. $\Delta |\mu|$\textsuperscript{\ref{DeltaDistFootNote}}) are shown for RRL stars around Pal 13 that passed our selection cuts (see Figure \ref{Pal13BeforeAfterCuts}). The CMD of Pal 13 is also shown. The blue cross corresponds to Pal 13 and the orange filled circles correspond to RRL stars inside Pal 13. The four stars shown in red represent stars that have escaped from Pal 13 and are assigned to Group A. Three of these stars (red-empty asterisks, Star $\#1$, $\#2$, and $\#3$) were previously identified by \citet{shipp2020} and we detect an additional star (red filled asterisk, Star $\#4$) that we believe may have escaped from Pal 13.}
\label{Pal13FullInfo}
\end{figure*}

\subsection{Pal 13} \label{sec:pal13}

Pal 13 is a faint (M$_{V}$ between $\sim$ $-$3.7 mag and $-$2.8 mag), low-mass, and low-luminosity GC located in the halo at a distance of $\sim$ 26 kpc. Multiple populations have been detected in Pal 13 and the implications of its mass-loss has been discussed \citep{tang2020}. 

Whether Pal 13 experiences tidal disruption has been debated repeatedly. Recently, \citet{Piatti2020} suggested the presence of stars escaping the cluster. As discussed in Section \ref{sec:intro}, \citet{shipp2020} identified tails around Pal 13 using a matched-filter technique based on stars at the
blue edge of the MS. In addition to that, they found signatures of tidal streams around Pal 13 by detecting six RRL stars of which three are located in its tail.

Since our aim is to find RRL stars around GCs, we start by applying our method and selection cuts to Pal 13 and then compare our findings with results from \citet{shipp2020}. We adopt the [Fe/H]$_{Pal13}$ and D$_{Pal13}$ values of Pal 13 listed in Table \ref{GCTableSummary} and we compute the corrected PM for the solar reflex (using the GALA package from \citealt{GalaPrice2017}) using the raw PM provided in the latter table. 





As shown in Figure \ref{Pal13BeforeAfterCuts}, there are seven RRL stars around Pal 13 (three cluster members and four located outside the cluster) that passed our selection cuts. Of these, six stars were previously identified by \citet{shipp2020}. We carefully examine the properties of these stars and compare them with the properties of Pal 13. 

Figure \ref{Pal13FullInfo} shows the (R.A. vs Dec.), ($\Delta$XY vs. $\Delta$Dist.), ($\Delta$$\mu_{\alpha}$ vs. $\Delta$$\mu_{\beta}$), and ($\Delta$$\theta$ vs. $\Delta |\mu|$) for Pal 13 and the seven RRL stars around Pal 13. In addition, we plot the (G$_{BP} - G_{RP}$ vs. G) CMD diagram of Pal 13 using Gaia EDR3 data.

$\Delta$XY is the spherical separation between the GC and each RRL star and is computed using Astropy\footnote{\url{https://docs.astropy.org/en/stable/coordinates/matchsep.html}} \citep{astropy:2013, astropy:2018}. $\Delta$Dist.\footnote{$\Delta Dist = D_{GC} - d_{RRL}$. Similarly for $\Delta$R.A., $\Delta$Dec.,  \\ $\Delta$$\mu_{\alpha}$, $\Delta$$\mu_{\beta}$, $\Delta$$\theta$, and $\Delta |\mu|$ \label{DeltaDistFootNote}} measures the difference in distances between the GC and the RRL stars and similarly for $\Delta$$\mu_{\alpha}$\textsuperscript{\ref{DeltaDistFootNote}}, $\Delta$$\mu_{\beta}$\textsuperscript{\ref{DeltaDistFootNote}}, $\Delta$$\theta$\textsuperscript{\ref{DeltaDistFootNote}}, and $\Delta |\mu|\textsuperscript{\ref{DeltaDistFootNote}}$.




\begin{figure*}
\centering
\hspace*{-0.5in}
   \includegraphics[scale = 0.46]{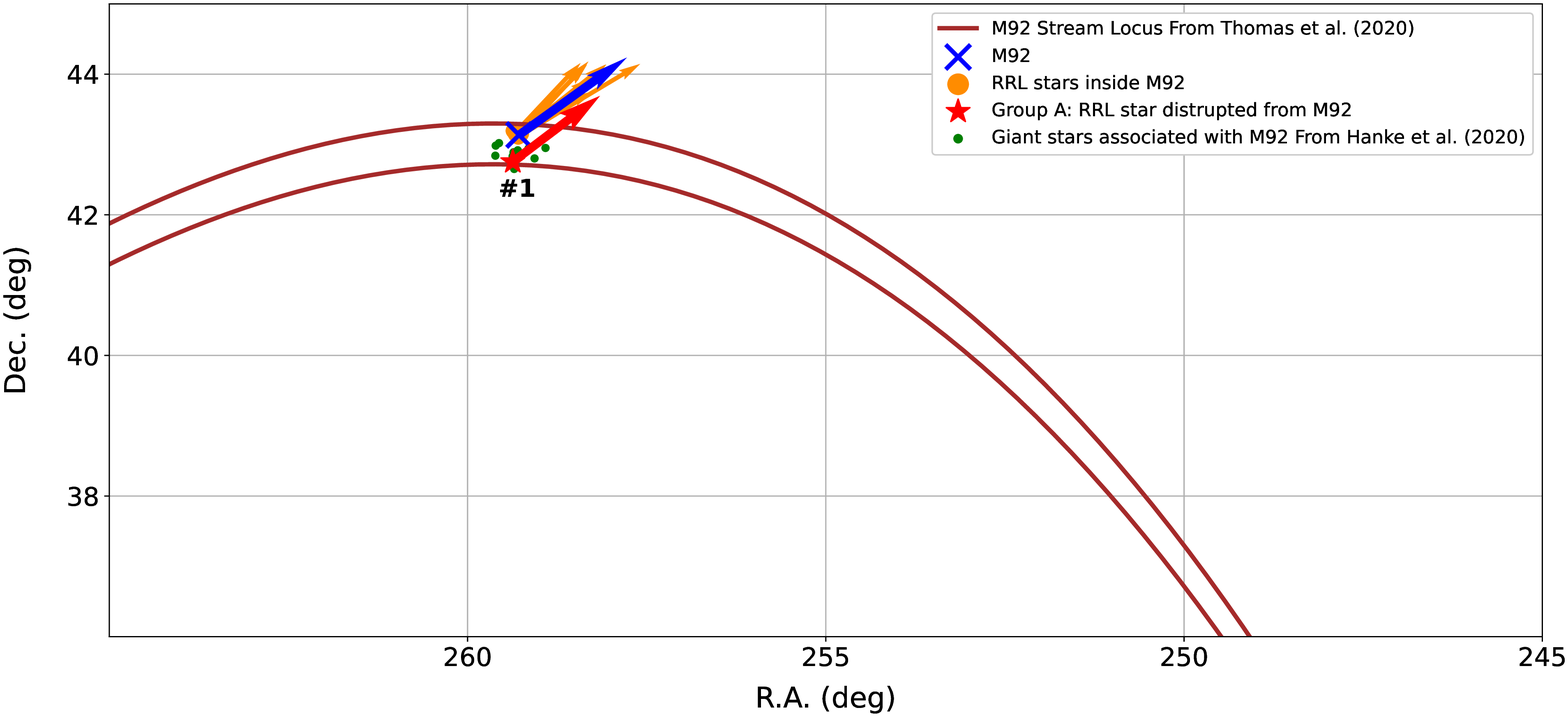}   
\caption{This plot shows eleven RRL stars that passed our selection cuts and we highlight the stream's location of M92 \citep{thomas2020} in brown lines. The location and PM of M92 are shown in blue and the ten RRL stars located inside M92 are shown as orange circles. Giant stars associated with M92 by \cite{hanke2020} are shown in green. We believe that Star $\#$1 represented as a red asterisk and located inside the stream path has escaped from M92 (Group A).}
\label{M92AfterCuts}
\end{figure*}

In Figure \ref{Pal13FullInfo}, the blue cross corresponds to Pal 13. The orange filled circles (3 stars) represent RRL stars inside Pal 13. The four red asterisks correspond to RRL stars that we detected outside Pal 13. Three out of these stars (Star $\#1$, $\#2$, and $\#3$) were previously detected by \citet{shipp2020} and an additional RRL star (red filled asterisk, Star $\#4$) was identified in our study. Green filled circles correspond to non-RRL stars belonging to Pal 13. 

We believe that the four stars shown in red in Figure \ref{Pal13FullInfo} have indeed escaped from Pal 13 (Group A stars) and provide evidence of its disruption. We present the properties and groups of these stars in Table \ref{TableEscapers}.

\begin{figure*}
\centering
\hspace*{-0.5in}
   \includegraphics[scale = 0.46]{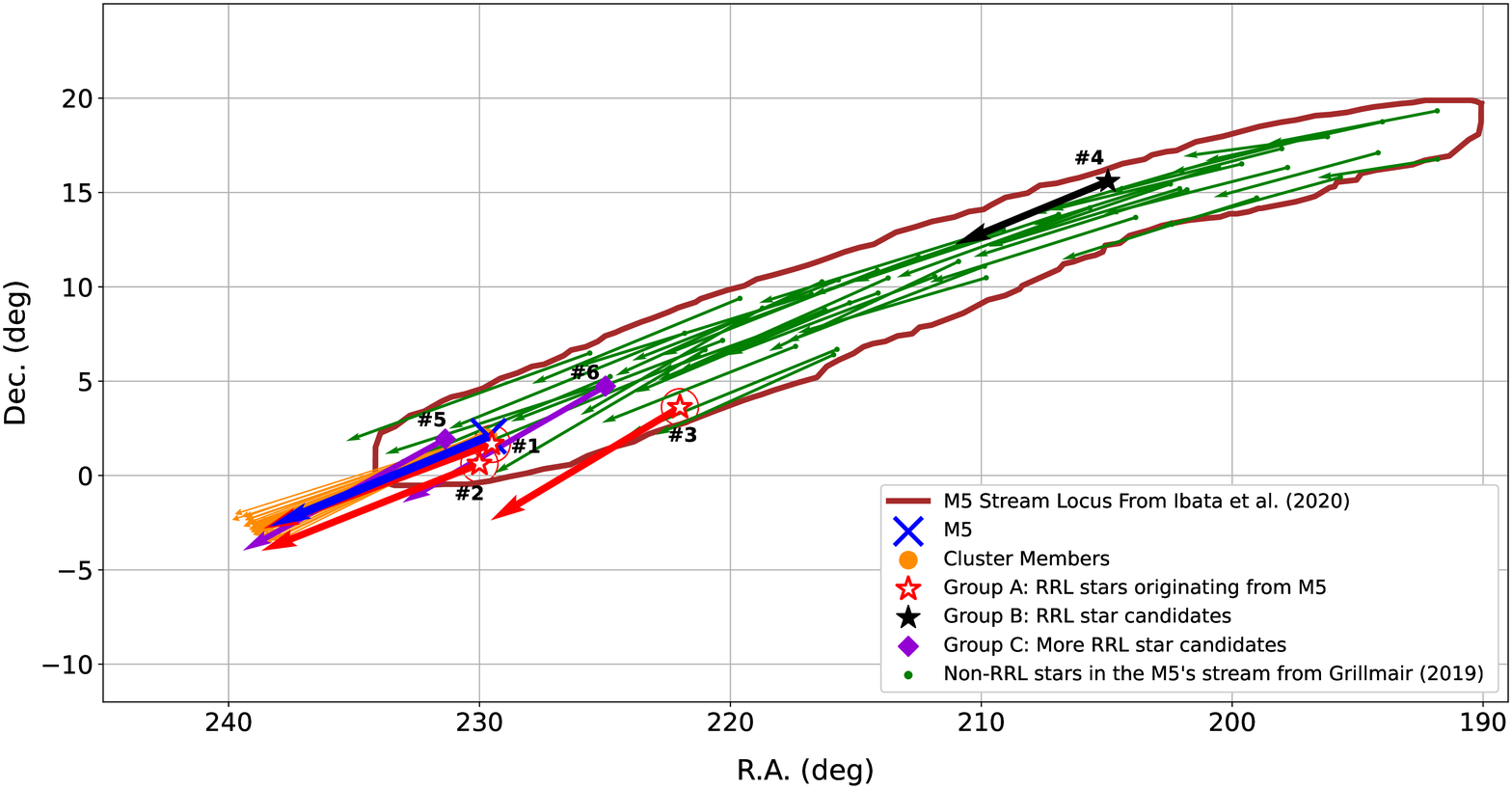}   
\caption{This plot shows the positions and PM arrows of six RRL stars and 50 non-RRL stars that might be associated with M5. The M5 stream position is adopted from \citet{ibata2020} and is shown in brown. Non-RRL stars from \citet{grillmair2019} that are associated with the M5 stream are plotted in green. Three Group A stars ($\#$1, $\#$2, and $\#$3), one Group B star ($\#$4), and two Group C stars ($\#$5, and $\#$6) are represented as red asterisks, black asterisks, and violet squares, respectively. RRL cluster members are shown in orange.}
\label{M5AfterCuts1}
\end{figure*}

\subsection{M92 (NGC 6341)} \label{sec:m92}

M92 is a bright and metal-poor GC located in the Hercules constellation. Extratidal stars were detected around M92 by \citet{jordi2010} and most recently, a stream was detected around M92 by \citet{sollima2020} and \citet{thomas2020}. Specifically, \citet{thomas2020} used BHB, MS, and RGB stars and found that the stream around M92 is $\sim$ 17$^{\circ}$ long and $\sim$ 0.29$^{\circ}$ wide and they provide a locus for its location.


There are 500 RRL stars spanning an area of $\sim$ 250  square degrees around M92 of which eleven stars passed our selection cuts. 

In Figure \ref{M92AfterCuts} we plot the locations of eleven stars that passed our cuts in addition to the locus of the stream (brown lines, adopted from \citealt{thomas2020}). We plot the locations of some of the giant stars that were identified by \citet{hanke2020} to be associated with M92 in green. 

Out of the eleven stars, ten are clusters members (orange circles) and one is located inside the stream locus and close to the cluster (red asterisk, Star $\#$1). We note that the association between Star $\#$1 and M92 has been previously suspected \citep{kundo2019}. Additionally, Star $\#$1 and more than five giant stars from \citet{hanke2020} seem to be clumped in the same region. After comparing the properties of Star $\#$1 with M92's, we believe with high confidence that Star $\#$1 has escaped from M92 so we assign it to Group A. 


\subsection{M5 (NGC 5904)} \label{sec:m5}

M5 is a nearby and bright GC located in the Serpens constellation. It is located at D$_{M5}$ = 7.5 $\pm$ 0.3 kpc \citep{ferro2016} and has a metallicity of [Fe/H]$_{M5}$ = $-$1.29 $\pm$ 0.05 dex.


\citet{jordi2010} found weak indications for a tidal stream emanating from M5. Subsequently, a 50$^\circ$ long stream was identified as a trailing tidal tail of M5 independently by \citet{grillmair2019} and \citet{ibata2020}. \citet{grillmair2019} provides the position of 50 stars that are most likely to be associated with M5 while \citet{ibata2020} plot the position of the stream.

We apply our method to search for RRL stars inside the M5 stream locus. We apply the selection cuts, analyze different 2D and 3D plots and find six stars that might have escaped from M5. The results are shown in Figure \ref{M5AfterCuts1}. In the latter figure, the position of the stream adopted from \citet{ibata2020} is represented in brown and the 50 non-RRL stars from \citet{grillmair2019} are shown in green.

Out of the six stars, three stars ($\#$1, $\#$2, and $\#$3, red asterisks in Figure \ref{M5AfterCuts1}) are most likely to have escaped from M5 (e.g., Group A), one star ($\#$4, black asterisk) has a lower probability to be associated with M5 (Group B), and an additional two stars ($\#$5 and $\#$6, violet squares) have less likelihood to be associated with M5 (Group C). See Table \ref{TableEscapers} for more information.

\subsection{NGC 5466} \label{sec:ngc5466}

Extended tidal tails around NGC 5466 were discovered by \citet{odenkirchen2004} and more of its stellar streams were detected and studied by \citet{belokurov2006,grillmair2006,weiss2018} and \citet{ibata2020}.

We manually trace the stream locus for NGC 5466 from \citet{ibata2020} and apply our selection cuts for stars inside or close to the stream locus. 

We find with that two RRL stars have escaped from NGC 5466 and these are represented as red asterisks in Figure \ref{FourGCs}a and are marked as Group A stars. We also suspect two additional stars (one in Group B and one in Group C) that might have been left behind by NGC 5466. Cluster members are shown in orange circles, NGC 5466 is represented as a blue cross, and the NGC 5466 stream locus from \citet{ibata2020} is plotted in brown in the same figure. 


\subsection{NGC 1261} \label{sec:ngc1261}

NGC 1261 is an old GC, located at a distance of D$_{NGC1261}$ = 17.2 $\pm$ 0.4 kpc and has a metallicity of [Fe/H]$_{NGC1261}$ = $-$1.42 $\pm$ 0.05 dex \citep{Ferro2019}. 


Candidate streams around NGC 1261 were detected but not confirmed in different studies. For instance, \citet{shipp2018} detected a stream around NGC 1261 by applying matched-filter techniques on data from the the Dark Energy Survey (DES; \citealt{DES2016}). \citet{ibata2020} detected a similar feature using the \emph{STREAMFINDER} algorithm \citep{malhan2018} on DR2 and EDR3 data.

\begin{figure*}
\centering
\hspace*{-0.4in}
   \includegraphics[scale=0.41]{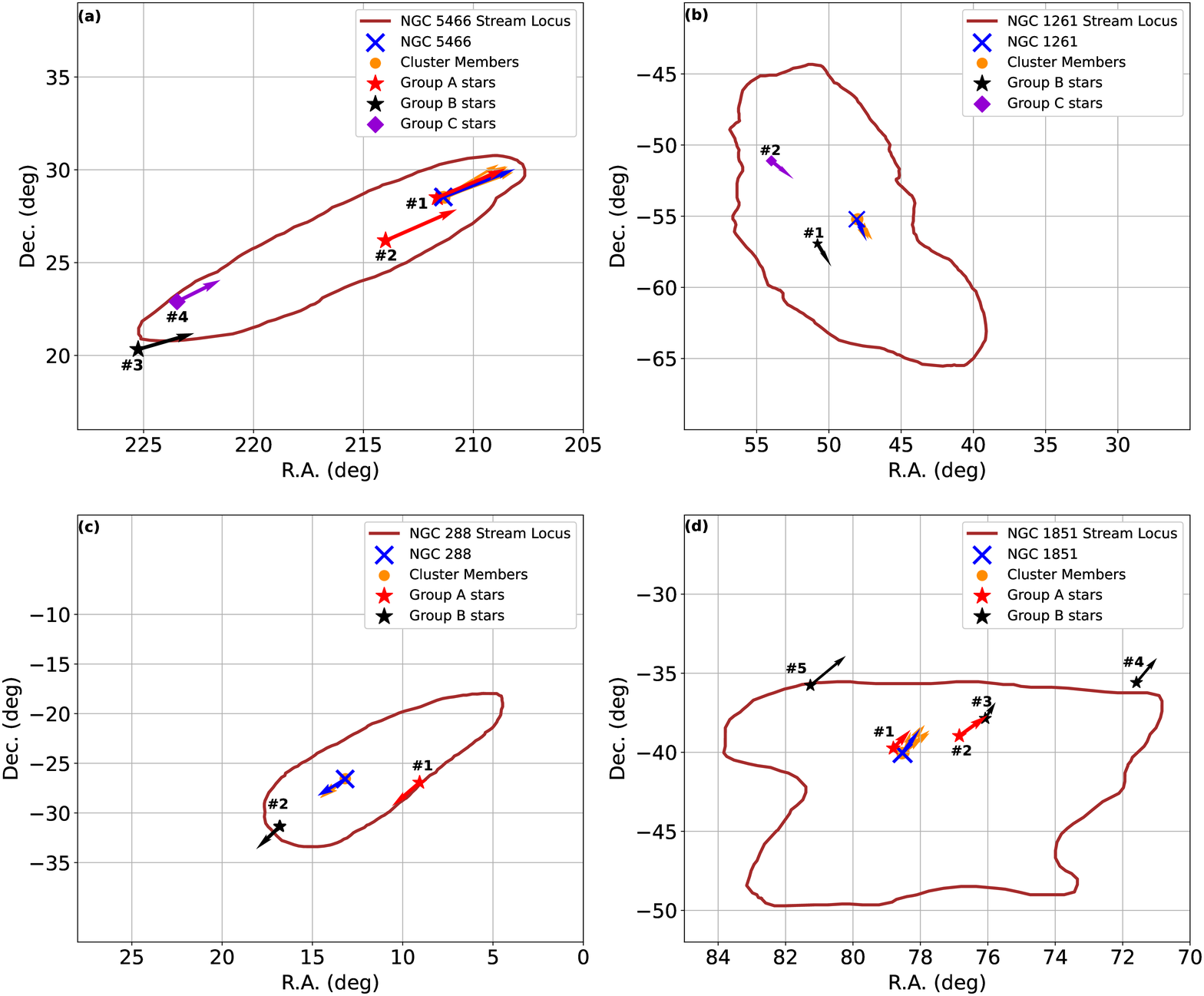}   
\caption{We present RRL candidates in panels (a), (b), (c), and (d) that might be associated with NGC 5466, NGC 1261, NGC 288, and NGC 1851, respectively. Brown lines represent the adopted stream loci from \citep{ibata2020}. Group A, B, and C stars are shown in red asterisks, black asterisk, and violet squares, respectively. Blue crosses correspond to the location of the GCs and orange circles represent cluster members. All arrows show the PMs of the relevant stars.}
\label{FourGCs}
\end{figure*}

Two candidate stars have passed our selection cuts (one Group B and one Group C). We plot the location of NGC 1261 in Figure \ref{FourGCs}b. The orange filled circles are cluster members, the black asterisk represents the RRL stars (Group B, $\#$1) while the violet square represents an RRL star that is close to the GC and inside the stream but with lower probability to have escaped from NGC 1261 ($\#$2) . 



\subsection{NGC 288} \label{sec:ngc288}

Different studies have detected possible tidal features at different distances associated with NGC 288 (e.g., \citealt{leon2000, grillmair2013}). Most recently, \citet{shipp2018} and \cite{ibata2020} presented more evidence about the properties and location of a stream associated with the latter cluster. We thus search for RRL stars located close to the position of the latter stream. 


In total, we find two stars that might be associated with NGC 288's stream and we plot them in Figure \ref{FourGCs}c. Star $\#$1 shown as red asterisk is most likely associated with the stream (e.g., Group A). An additional RRL star passed our cuts and is shown as a black asterisk ($\#$2, Group B).


\subsection{NGC 1851} \label{sec:ngc1851}

Different studies attempted to look for features associated with NGC 1851 but the results were not always consistent (e.g., see \citealt{olszewski2009,carballo2018,shipp2018,stringer2020,ibata2020}).

Our search reveals five RRL candidates that might be related to NGC 1851's stream that we show in Figure \ref{FourGCs}d. Specifically, we find two Group A stars ($\#$1 and $\#$2) and three Group B stars ($\#$3, $\#$4, and $\#$5).



\subsection{Gaia Parallaxes} \label{sec:parallax}

As discussed in Section \ref{sec:distances}, distances (d$_{RRL}$) presented in Table \ref{TableEscapers} were calculated using the $M_{G}$ -- [Fe/H] and $M_{V}$ -- [Fe/H] relations. For comparison purposes, we re-calculate distances to the latter stars using Gaia EDR3 parallaxes, hereafter d$_{GAIA}$. We select stars with low uncertainty parallax values: $\sigma_{\bar{\omega}}$/${\bar{\omega}}$ $\textless$ 0.1 where $\bar{\omega}$ and $\sigma_{\bar{\omega}}$ are the EDR3 parallaxes and their uncertainties in mas, respectively. The parallax values were zero point corrected using the \citet{lindegren2020} solution\footnote{\url{https://gitlab.com/icc-ub/public/gaiadr3_zeropoint}\label{zeropoint}}.

\begin{figure}
\centering
\hspace*{-0.3in}
\includegraphics[scale=0.27]{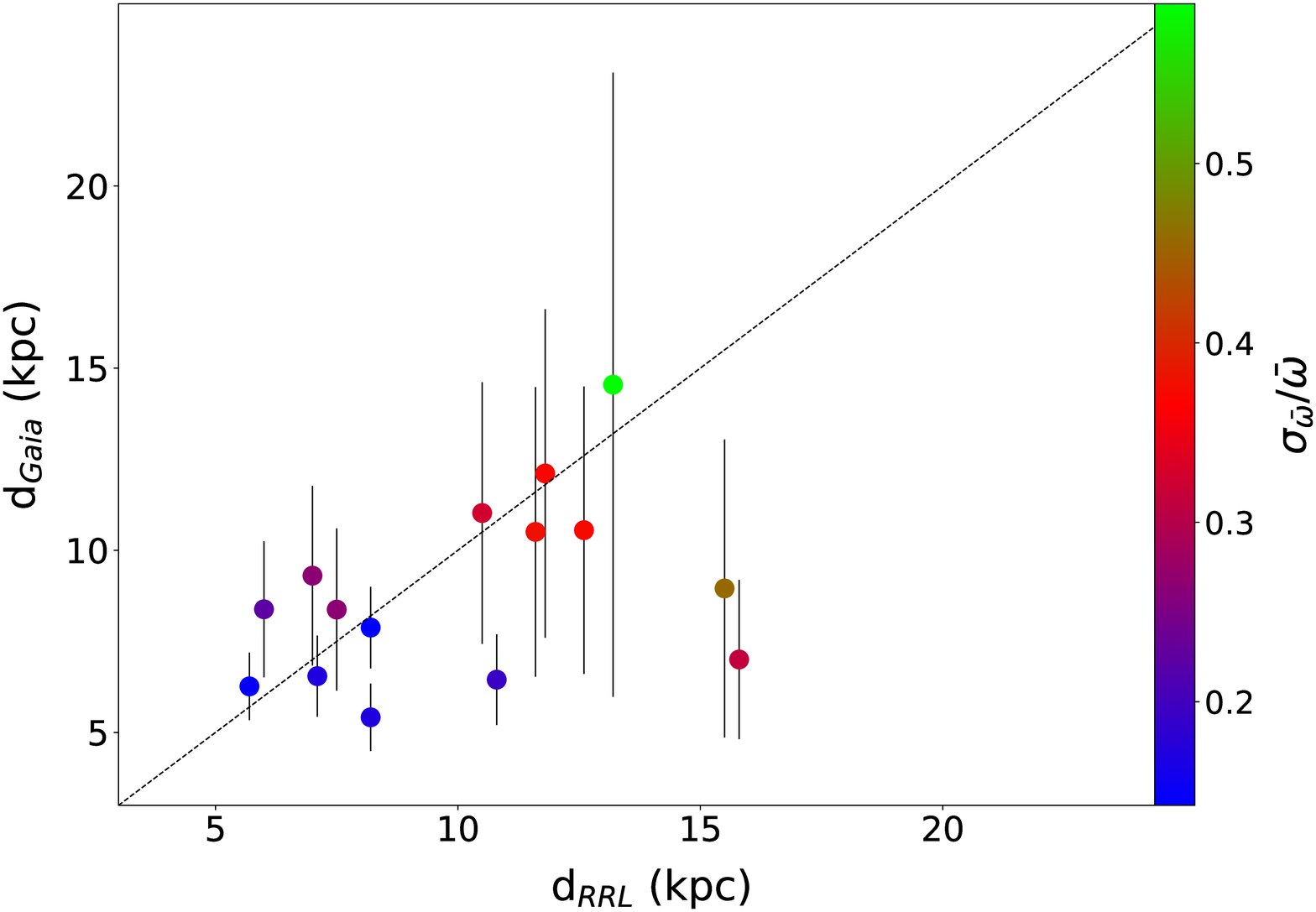}
\caption{We compare distances of RRL stars derived from the $M_{G}$ -- [Fe/H] and $M_{V}$ -- [Fe/H] relations (d$_{RRL}$) and from the Gaia EDR3 parallaxes (d$_{GAIA}$) for stars with $\sigma_{\bar{\omega}}$/${\bar{\omega}}$ $\textless$ 0.1. The colors of the data points reflect the corrected $\sigma_{\bar{\omega}}$/${\bar{\omega}}$ values of each star according to the legend. Distances derived from both methods are in good agreement and the scatter from the relation ranges between 1 kpc and 3 kpc.}
\label{distdistance_comparison}
\end{figure}

From the 24 RRL stars found in our table, 15 stars have $\sigma_{\bar{\omega}}$/${\bar{\omega}}$ $\textless$ 0.1. We show their d$_{GAIA}$ vs. d$_{RRL}$ scatter plot in Figure \ref{distdistance_comparison} where the corrected $\sigma_{\bar{\omega}}$/${\bar{\omega}}$ values are color coded according to the legend. The scatter from the relation ranges between 1 kpc and 3 kpc and is minimal for stars with smaller $\sigma_{\bar{\omega}}$/${\bar{\omega}}$ values. This increases the reliability of the method we are using in our study to find the distances of RRL stars in the aim of detecting possible escapers.

\section{Summary} \label{sec:summary}

The aim of this study is to identify RRL stars that escaped from seven GCs. To do that, we use a sample of $\sim$ 150,000 RRL stars by cross-matching the catalogs of RRL stars from CRTS \citep{drake2013a} and Gaia \citep{clementini2019} with each other and with the Gaia EDR3 database. Properties of the GCs under investigation are listed in Table \ref{GCTableSummary} and their PMs are adopted from \cite{eugene2019}. The seven GCs were selected as their stream loci were previously identified and because they fall in the footprint of the data and surveys we are using in this study. A comprehensive analysis on other GCs will be provided in a companion paper (Abbas et al. 2021, in preparation).

We study the PMs, distances, and other characteristics of the RRL stars in the vicinity of the selected GCs and we compare them with the properties of the GCs. 
We use the $M_{G}$ -- [Fe/H] \citep{tatiana2018} and $M_{V}$ -- [Fe/H] \citep{drake2013a} absolute magnitude-metallicity relations to find distances and rely on the locations of previously identified streams from different studies to constrain our search (e.g., \citealt{grillmair2019,ibata2020, thomas2020, shipp2020}).

We find several RRL stars that might have escaped from different GCs and show them in Figures \ref{Pal13FullInfo}, \ref{M92AfterCuts}, \ref{M5AfterCuts1} and \ref{FourGCs}. Confirming whether an RRL star really escaped from a GC can be very challenging. We thus assign each RRL star to Group A, B, or C. Group A stars are most likely to have been erupted from the GC. Group B stars have intermediate likelihoods while Group C stars have less likelihoods. 

In total, we identify 24 field RRL stars that might be associated with seven GCs. Specifically, we associated four field RRL stars with Pal 13 (Group A), one star with M92 (Group A), six stars with M5 (three Group A, one Group B, and two Group C), four stars with NGC 5466 (two Group A, one Group B, and one Group C), and five stars with NGC 1851 (two Group A and three Group B). We also report two distinct stars that likely originated from each of the two following GCs: NGC 1261 (one Group B and one Group C), and NGC 288 (one Group A and one Group B). We list all our findings in Table \ref{TableEscapers}.

We visually inspected the phased light curves of each RRL star in the latter table using periods from Gaia or CRTS (or both when available). Based on the periods, phased light curves, and on the position of these stars on the CMD diagrams, we strongly believe that these are indeed RRL stars and their classifications are highly reliable. This results in a high purity levels and supports our analysis.

At the same time, given that RRL stars are comparatively scarce objects and that the combined catalog of RRL stars we are using isn't 100$\%$ complete, and since we have applied relatively conservative cuts (i.e., Equations \ref{cut_dist}, \ref{cut_length}, and \ref{cut_angle}), there could be additional RRL stars that we missed that might have escaped from our studied GCs.


Identifying RRL stars that escaped form a GC doesn't essentially infer the disruption of the GC. However, if the extratidal RRL stars fall in tidal tails that were previously identified using other type of stars, this could be a comprehensive evidence for the disruption of the GC. In all cases this can help us better understand the shape of the Galactic potential and the history of evolution of the Galaxy. Finding additional RRL stars in the surrounding fields of other GCs would be ideal for a broader understanding about the internal and external dynamical evolution of GCs.

\acknowledgments

The authors would like to thank the anonymous referee for a prompt report and useful comments. This material is based upon work supported by Tamkeen under the NYU Abu Dhabi Research Institute grant CAP$^3$. M.A. and E.K.G. acknowledge support by the Collaborative Research Center ``The Milky Way System" (SFB 881, Project-ID 138713538, subproject A03) of the German Research Foundation (DFG).


\begin{table*}
 \centering
  \caption{The Escapers: RRL Stars Originating from GCs}
  \begin{tabular}{cccccccccccccc}
  \hline
  \hline
  G.C. & N$_{RRL}$\footnote{Number of RRL Candidates in each GC.} & Star $\#$\footnote{Star number based on Figures \ref{Pal13FullInfo}, \ref{M92AfterCuts}, \ref{M5AfterCuts1}, and \ref{FourGCs}.}  & R.A.\footnote{Equatorial J2000.0 R.A. and Dec. are given in decimal degrees.\label{RaDec}} &
  
  Dec.\textsuperscript{\ref{RaDec}} & 
  
  d$_{RRL}$ \footnote{Distances to the RRL stars in kpc.} &
  
  P$_{V}$(d)\footnote{\smash{Periods and amplitudes based on the CRTS RRL stars catalog \citep{drake2013a}.\label{PerAmpsCRTS}}} &

  P$_{G}$(d)\footnote{\smash{Periods and amplitudes based on the Gaia RRL stars catalog \citep{clementini2019}.\label{PerAmpsGaia}}} &
  
  A$_{V}$\textsuperscript{\ref{PerAmpsCRTS}} & A$_{G}$\textsuperscript{\ref{PerAmpsGaia}} & A$_{BP}$\textsuperscript{\ref{PerAmpsGaia}} & A$_{RP}$\textsuperscript{\ref{PerAmpsGaia}} &

  Group\footnote{{The Group reflects the likelihood of each star to have escaped from the assigned GC. Group A stars have very high likelihoods while Group B and Group C stars have intermediate and low likelihoods, respectively. }}  
  
  & 
  Catalog
  \\
  \hline \hline
  Pal 13 & 4 &  $\#$1  & 341.484 & 5.492 & 20.4 & 0.581 & NA & 0.68 & NA & NA & NA & A & CRTS \\

    &  &  $\#$2  & 343.878 & 7.820 & 21.9 & 0.555 & 0.555 & 0.78 & 0.623 & NA & NA & A & Both\\ 
      
     &  &  $\#$3  & 347.372 & 14.154 & 23.9 & NA & 0.609 & NA & 0.457 & NA & NA & A & Gaia\\

     &  &  $\#$4  & 345.559 & 13.961 & 24.7 & 0.581 & 0.595 & 0.5 & 0.404 & NA & NA & A & Both\\ 
     \hline
   
   M92 & 1 & $\#$1  & 259.368 & 42.754 & 8.2 & 0.728 & 0.728 & 0.47 & 0.552 & 0.591 & 0.313 & A & Both \\ 

     \hline

  M5 & 6 & $\#$1  & 229.509 & 1.677 & 7.0 & 0.650 & NA & 0.2 & NA & NA & NA & A & CRTS\\
  
   & & $\#$2  & 230.001 & 0.618 & 7.5 & 0.616 & NA & 0.42 & NA & NA & NA & A & CRTS\\
   
   &  & $\#$3  & 222.01 & 3.620 & 8.2 & 0.470 & NA & 0.93 & NA & NA & NA & A & CRTS\\
   
   &  & $\#$4  & 204.948 & 15.592 & 7.1 & 0.532 & 0.532 & 1.09 & 1.09 & 1.04 & 0.76 & B & Both\\
   
  & & $\#$5  & 231.364 & 1.933 & 6.0 & 0.516 & NA & 1.01 & NA & NA & NA & C & CRTS\\
  
  & & $\#$6  & 224.978 & 4.740 & 5.7 & 0.352 & NA & 0.35 & NA & NA & NA & C & CRTS\\
  
\hline

 NGC 5466 & 4 & $\#$1  & 211.637 & 28.507 & 15.5 & 0.577 & 0.577 & 1.05 & 1.02 & 0.854 & 0.533 & A & Both\\
   
     &  & $\#$2  & 213.992 & 26.186 & 14.9 & 0.623 & 0.623 & 1.02 & 1.112 &  NA & NA & A & Both\\
  
     &  & $\#$3  & 225.248 & 20.325 & 15.1 & 0.61 & 0.61 & 0.63 & 0.564 & 0.574 & 0.286 & B & Both\\
     
  & & $\#$4  & 223.478 & 22.896 & 13.2 & 0.669 & 0.669 & 0.41 & 0.417 & 0.524 & 0.316 & C & Both\\
  

     \hline

           NGC 1261    & 2  &  $\#$1  & 50.789 & -56.925 & 13.7 & NA & 0.478 & NA & 1.0 & 1.03 & 0.58\ & B & Gaia \\

    &  &  $\#$2  & 53.981 & -51.11 & 15.8 & NA & 0.472 & NA &  0.886 & 0.446 & 0.482 & C & Gaia \\




   \hline

   NGC 288 & 2 & $\#$1  & 9.064 & -26.93 & 8.6 & 0.717 & 0.717 & 0.77 & 0.755 & 0.903 & 0.431 & A & Both\\
   
   &  & $\#$2  & 16.806 & -31.35 & 9.8 & 0.613 & 0.613 & 0.58 & 0.592 & 0.781 & 0.434 & B & Both\\
   
   
  
    \hline
  
  \hline
  NGC 1851 & 5 & $\#$1  &  78.802 & -39.745 & 11.8 & 0.59 & NA & 0.74 & NA & NA & NA & A & CRTS\\
  
  & & $\#$2  &  76.845 & -38.962 & 11.6 & 0.58 & 0.58 & 0.56 & 0.708 & 0.632 & 0.5 & A & Both\\
  
  & & $\#$3  & 76.075 & -37.859 & 10.8 &  NA & NA & NA & 0.355 & 0.399 & 0.262 & B & Gaia\\
  
  & & $\#$4  &  71.591 & -35.59 & 12.6 & 0.61 & 0.61 & 0.92 & 1.0 & 0.87 & 0.58 & B & Both\\
  
  
  & & $\#$5  & 81.264 & -35.77 & 10.5 & 0.797 & 0.797 & 0.51 & 0.546 & 0.724 & 0.398 & B & Both\\
  
    \hline

  \hline
\end{tabular}
\label{TableEscapers}
\end{table*}

\newpage

\bibliography{sample63}{}
\bibliographystyle{aasjournal}

\end{document}